\documentclass[aps,prl,twocolumn,floatfix,showpacs,groupedaddress]{revtex4}
\usepackage{graphicx}
\usepackage{dcolumn}
\usepackage{bm}
\usepackage{amssymb}
\begin{document}
\title{Soft Elasticity in Nematic Liquid-Crystal Networks.}

\author{Simiso K. Mkhonta,$^1$ Dan Vernon,$^1$ K. R. Elder,$^2$
and Martin Grant$^1$
}
\date{\today}

\affiliation{$^1$Physics Department,Rutherford Building, 3600 rue University,
McGill University, Montr\'eal, Qu\'ebec, Canada H3A 2T8 \\
$^2$Department of Physics, Oakland, Rochester, Michigan 48309-4487, USA}

\begin{abstract}
Liquid-crystal networks consist of weakly crosslinked polymers that are
coupled
to liquid-crystal molecules. The resultant hybrid system has rich elastic
 properties.
We develop a phase field model to describe mechanical properties of
a hexagonal liquid-crystal network.  The hexagonal liquid-crystal network is
found to have soft shear deformations. The elastic properties are predicted
analytically
and confirmed with numerical simulations.  In addition the model naturally
 incorporates
non-linear elasticity and dislocations or disclinations.
\end{abstract}
\pacs{62.20.-x, 83.80.Va, 61.30.Dk}
\maketitle

Soft elasticity is an exotic elastic phenomenon
 found in liquid-crystal networks which
incorporates liquid-crystalline order and rubber
 elasticity~\cite{Warner06}. The simplest liquid-crystal network (LCN)
 consists of weakly crosslinked polymers with the constituent rodlike
liquid-crystal molecules
 in the nematic phase. A nematic LCN can relieve a shear stress via
the additional degrees of freedom that arise from the constituent nematic
molecules.  In addition to exhibiting rich elastic behavior, LCNs 
are being considered in new
 technological applications that include artificial
muscles~\cite{Shenoy02}, opto-mechanical actuators~\cite{Camacho04}, and
tunable mirror-less lasers~\cite{Finkelmann01}.

Ideally, the nematic molecules are free to wiggle about their
anisotropic axes.  As a consequence, when a
shear deformation is applied to a nematic LCN in a plane that involves
 the nematic director, the nematic molecules can tilt locally
 such that the system retains
its equilibrium configuration.
This phenomena is referred to as soft elasticity~\cite{Warner06} and
is in the same spirit as the prediction of Golubovic-Lubensky that
an anisotropic glass that breaks spontaneously a continuous
 symmetry
must have a vanishing shear modulus~\cite{Golubovic89}.

In this article we show that the phenomenon of soft elasticity
 is not limited to nematic gels ~\cite{Lubensky02}
and smectic gels~\cite{Stenull05} but can also be realized in
 networks with more translational order.
 Our system is distinct from the ``soft crystal" phases of a liquid crystal
fluid~\cite{deGennes95} due to the topological constraints 
offered by the network. Our hybrid system thus shrinks or 
elongates spontaneously depending on the orientational
 distribution of the 
nematic molecules.
Our model could be tested experimentally in a block copolymer system
that can self-assemble in a liquid-crystalline environment into an 
ordered network. For example, a \emph{ABA} triblock copolymer solution 
has been shown to physically crosslink when the \emph{A}-type monomers 
are phobic to the liquid crystal solvent~\cite{Kornfield04}.

To capture the properties of a LCN we couple
the network molecular shape to the orientational order of the background
nematic molecules. The coupling is such that in the nematic
phase, the network is elongated in a direction determined 
by the average orientation of the nematic molecules. 
In this state the 
network density field $\psi(\mathbf{r})$ is anisotropic.  We assume that the network molecular lengthscale that is 
obtained from its characteristic 
wavenumber $q_0$ is much larger than the dimension of the nematic 
molecules $a$. 
This assumption holds in the case of the $\emph{ABA}$ triblock network 
system  where 
the radius of gyration of the polymers $R_g \cong 1/q_0 >> a$.
This allows us to utilise the nematic director field
$\theta(\mathbf{r})$ to describe the nematic molecules.

Under these conditions we propose the phenomenological free energy functional
\begin{eqnarray}
F &=& \int d^2\mathbf{r}\,\{K\left(\nabla \theta \right)^2/2
+ \tau\psi^2/2 +\lambda\psi^3/3+ \psi^4/4 +\nonumber\\
&& \left[q^2_0\psi+A\partial
_{xx}\psi+B\partial_{yy}\psi +C\partial_{xy}\psi\right]^2/2\},
\label{eqn:energyd}
\end{eqnarray}
where
$A(\theta) = \cos^2 \theta + \kappa^2 \sin^2 \theta$,
$B(\theta)= \sin^2 \theta + \kappa^2 \cos^2 \theta$, and
$C(\theta) = (\kappa^2-1)\sin {2\theta}$ and where $K$ is the Frank elastic
constant for
liquid crystals, $\kappa$ is the anisotropy ratio 
of the density fluctuations in the nematic state, 
$\tau$ is a control parameter and $\lambda < 0$ is a phenomenological  constant.
In the limit $\kappa =1$, the network fluctuations are 
isotropic regardless of the nematic order 
and thus the two fields are uncoupled.  

The  dynamics of the model are driven
by the minimization of the free energy, i.e.,
\begin{eqnarray}
\partial \psi(\mathbf{r},t)/\partial t
& =&\nabla^2 \left[\delta F/\delta \psi(\mathbf{r})\right], \nonumber\\
\partial \theta(\mathbf{r},t)/\partial t
&=&-\left[\delta F/\delta \theta(\mathbf{r})\right] +
 \mu\,\eta(\mathbf{r},t),
\label{eqn:dynamics}
\end{eqnarray}
where $\eta(\mathbf{r},t)$ is a noise field and $\mu$ is
the intensity of the noise.
For simplicity, the thermal fluctuations are only incorporated
in the orientational field such that $\eta$ is assumed to be a Gaussian
random function, with a mean variance
$\langle\eta(\mathbf{r},t),\eta(\mathbf{r'},t')\rangle = \delta
(\mathbf{r}-\mathbf{r'})\delta (t-t')$.
The square of the intensity of the noise is
linearly proportional to temperature.

The equilibrium states for Eq.~(\ref{eqn:energyd}) can be determined 
in mean field theory by considering the
minima of $F$. 
We can assume that the average
molecular orientation is along the y-axis [i.e., $\langle 
\theta(\mathbf{r})\rangle = 0$].
In this instance the free energy functional simplifies to
\begin{eqnarray}
F^{m} & \approx &\int d^2\mathbf{r}\,\{(r^2_{\perp}\partial_{xx}\psi
+r_{\parallel}^2\partial_{yy}\psi+q^2_0\psi)^2/2\nonumber\\
&&+\tau\psi^2/2 + \lambda\psi^3/3+ \psi^4/4 \},
\label{eqn:APFC}
\end{eqnarray}
where  $r^2_{\parallel} \equiv \langle B(\theta)\rangle$,
$r^2_{\perp} \equiv \langle A(\theta)\rangle$ and
the brackets imply an average over the
liquid crystal orientational distribution.
Since the nematic director is assumed to be along the $y-$coordinate,
from Eq.~(\ref{eqn:APFC}) we obtain that $r_{\parallel}$ and $ r_{\perp}$
define the periodicity of the network parallel and perpendicular to the
nematic director respectively.

The averaged free energy $F^{m}$ is in the form
of the Landau-Brazovskii (LB) 
phenomenological theory~\cite{Brazovskii75} that describes phase
transitions of a uniform system to a periodic state. The only difference
 is the scaling factor between the two coordinates.
The mean field phase diagram of this model is 
well known~\cite{Wickham03,Elder04}.
The approximate equilibrium solutions of the density
fluctuations are then similar to the crystalline solutions of the LB
theory. For example the hexagonal phase is described by
\begin{eqnarray}
\psi(x,y) &=& A_t \Big\{\cos\left[\sqrt{3} q_0x/
(2r_{\perp})\right]\cos\left[q_0y/(2r_{\parallel})\right]
 \nonumber \\
&& +\cos \left({q_0y}/{r_{\parallel}}\right)/2\Big\},
\label{eqn:opA}
\end{eqnarray}
where $A_t=4(-\lambda+\sqrt{-15\tau +\lambda^2})/15$ is
 the amplitude of the local density fluctuations. 
The sinusoidal solution Eq.~(\ref{eqn:opA}) is the leading term approximation of
 a Fourier series and is only valid for $0>\tau >> -1$. 

\begin{figure}
 \resizebox{80mm}{!}{\includegraphics{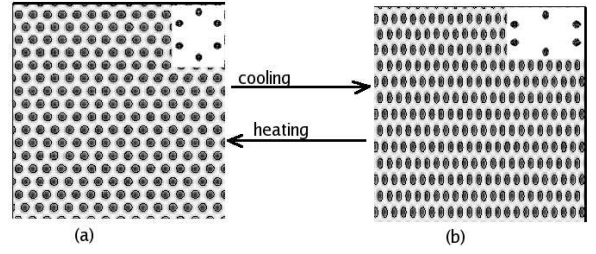}}
 \caption{\label{fig:ucrystal} Equilibrium configuration of the local
network density $\psi(\mathbf{r})$ with anisotropy $\kappa =2$: 
(a) in the isotropic phase ($\mu=20$) and  (b) in the nematic phase ($\mu=1$). The dark regions correspond to $\psi >0$ and  the light regions
correspond to $\psi <0$.  (Inset: structure factor)}
\end{figure}

At low temperatures, the nematic molecules 
have a preferred direction [$\theta(\mathbf{r})$ =0, in our case]
and thus we have $r^{nem}_{\parallel}=\kappa $ and $r^{nem}_{\perp} = 1 .
\label{eqn:nem}$
In the high-temperature isotropic phase, the nematic molecules have
 no preferred direction, i.e., $\theta(\mathbf{r})$ is a random
 field. Thus we have $r^{iso}_{\parallel}=r^{iso}_{\perp} =\sqrt {(\kappa^2 +1)/2}$.
The change of $ r_{\parallel}$ and $ r_{\perp}$ with temperature highlights
 elongations of the network lattice due to nematic ordering. An example of
 the network lattice distortion is presented in Fig.\ref{fig:ucrystal}. 
Spontaneous elongation of the network
during a isotropic-nematic transition is a hallmark of LCNs~\cite{Warner06}.

The continuum elastic free energy of our elongated hexagonal LCN
is constructed from the lattice symmetry of the network and the
coupling of its rotational deformation to the nematic
director orientation.  It is given by 
\begin{eqnarray}
E_{el}  &=&  \left[C_{11}(u_{xx}^2 +u_{yy}^2)
+2C_{12}u_{xx}u_{yy} +4C_{66}u_{xy}^2\right]/2\nonumber\\
&+& D_1 \left(\theta-\omega_{xy}\right)^2/2
+D_2\left(\theta-\omega_{xy}\right)u_{xy}\nonumber\\
&+& K\left(\nabla\theta \right)^2/2,
\label{eqn:elastic}
\end{eqnarray}
where $u_{\alpha\beta}=(\partial_\beta u_\alpha
+\partial_\alpha u_\beta)/2$ is the symmetric strain tensor and $\omega_{\alpha\beta}=(\partial_\beta u_\alpha
-\partial_\alpha u_\beta)/2$ is
the antisymmetric strain tensor. $u_{\alpha\beta}$ 
and $\omega_{\alpha\beta}$ describe
respectively the relative translations and rotations of the network due
to an applied deformation.
The first line in Eq.~(\ref{eqn:elastic}) is the elastic free energy of a
two dimensional hexagonal crystal~\cite{Landau86} and the second line is the
de Gennes elastic energy~\cite{deGennes82}, which describes the coupling
of the nematic molecules to local fluctuations of the network strands.

We will determine the elastic moduli from our molecular
free energy.  We will assume that the distortions are small,
so changes of the density amplitude $\psi(x,y)$ are negligible.
We will choose the molecular axis
and nematic director to be along the y-axis, where $\theta(\mathbf{r}) =0$, and
thus employ Eq.~(\ref{eqn:APFC}) as the effective free energy density.
\begin{figure*}
\resizebox{160mm}{!}{\includegraphics{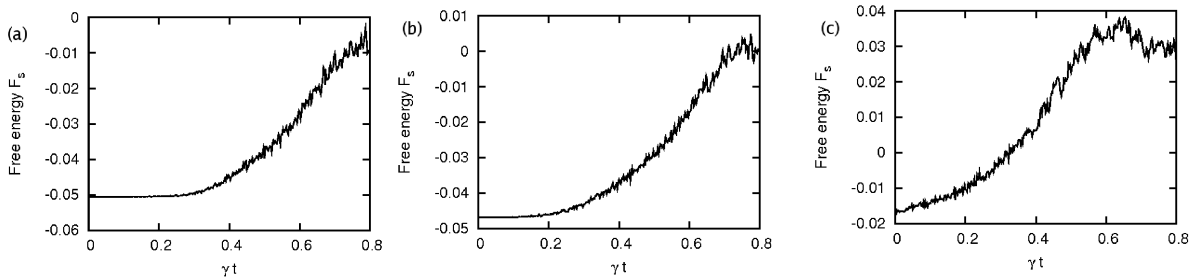}}
\caption{\label{fig:soft1} The free-energy density evolution after
commencement of a shear flow with a shear rate $\gamma = 0.002$.
The system is shown at noise strengths: $\mu=1.0$, $\mu=2.0$,
and $\mu=5.0$
 for (a)-(c) respectively. The molecular ratio is $\kappa =2$.}
\end{figure*}
A pure shear distortion is described by the density
$\psi= \psi(x+\zeta y, y+\zeta x)$, where $\zeta$ is the elastic strain.
In this state the shear free elastic energy per unit area
($f_{shear}\equiv [F^m(\zeta)-F^m(\zeta=0)]/{\rm area}$) can be
calculated by substituting a one mode approximation for $\psi$
[i.e., Eq.~(\ref{eqn:opA})] into Eq.~(\ref{eqn:APFC})  to obtain
\begin{equation}
 f_{shear}= \frac{3}{32}\left(\frac{r_{\parallel}}{r_{\perp}}
+ \frac{r_{\perp}}{r_{\parallel}}\right)^2q^4_0A_t^2\zeta^2
+{\cal O}(\zeta^4).
\label{eqn:pureshear}
\end{equation}
Eq.~(\ref{eqn:elastic}) can now be used to calculate
$C_{66}$ recalling that for a pure shear $u_{\alpha\beta}=0$, except
for $u_{xy}=\zeta$.  This gives
\begin{equation}
C_{66} =\frac{1}{4}\frac{\partial ^2f_{shear}}{\partial \zeta
  ^2}=\frac{3}{64}\left(\frac{r_{\parallel}}{r_{\perp}}
+ \frac{r_{\perp}}{r_{\parallel}}\right)^2A_t^2q^4_0 .
\label{eqn:C66}
\end{equation}
At low temperatures we have $r^{nem}_{\parallel}/r^{nem}_{\perp}=\kappa$, 
which means the shear modulus is strongly
 dependent on the network molecular anisotropy. Under a similar procedure,
 other physical elastic moduli such as the bulk modulus
$B_u =3A_t^2q^4_0/32$ and the deviatoric modulus $C_{d}= 3A_t^2q^4_0/16$
are obtained. These do not depend on the molecular aspect
ratio~\cite{Elder04} and note that when $r_{\parallel} = r_{\perp}$  
 then $C_{d}=C_{66}$ as expected for a hexagonal lattice.

A pure rotational deformation is described by
the strain tensor $\omega_{xy} = \zeta$ and $ u_{\alpha\beta} =0$.
The elastic energy terms in Eq.~(\ref{eqn:elastic}) all vanish
 except the term with coefficient $D_1$. In this state,
$\psi = \psi(x+\zeta y,y-\zeta x)$.
The evaluated free-energy density of the deformed state gives
\begin{equation}
D_1 =\frac{3}{16} \left( \frac{r_{\parallel}}{r_{\perp}}
-\frac{r_{\perp}}{r_{\parallel}}\right)^2 A_t^2 q^4_0.
\label{eqn:D1}
\end{equation}
From the difference between two simple shears along each coordinate 
we can determine $D_2$:
\begin{equation}
D_2 =\frac{\partial ^2f^y_{shear}}{\partial \zeta
  ^2} -\frac{\partial ^2 f^x_{shear}}{\partial \zeta
^2} = \frac{3}{16}\left( \frac{r^2_{\perp}}{r^2_{\parallel}}
-\frac{r_{\parallel}^2}{r^2_{\perp }} \right)A_t^2q^4_0.
\label{eqn:D2}
\end{equation}
The coupling moduli $D_1$ and $D_2$ vanish when the network density fluctuations are isotropic (where $r_{\perp} =r_{\parallel}$), i.e, in the isotropic liquid-crystal phase or when $\kappa = 1$.
It is interesting to note that these calculations for the molecular
moduli $C_{66}$, $D_1$ and $D_2$ are consistent with the
theoretical result of Warner and Terentjev \cite{Warner06} for
nematic elastomers derived from classical rubber elasticity.

When the nematic director is allowed to relax to an optimum state,
 then $\partial E_{el}/\partial \theta =0$ and
thus Eq.~(\ref{eqn:elastic}) reduces to
\begin{eqnarray}
E^R_{el} &=& C_{11}\left(u_{xx}^2+
u_{yy}^2\right)/2+C_{12}u_{xx}u_{yy}
+2\widetilde{C}_{66}u_{xy}^2\nonumber\\
& +& KD_2^2\left(\nabla
u_{xy}\right)^2/(2D^2_1)
+ K\left(\nabla \omega_{xy}\right)^2/2,
\label{eqn:soft}
\end{eqnarray}
where $\widetilde{C}_{66} = C_{66} -D_2^2/(4D_1)$ is
the renormalized shear modulus.

From Eqs.~(\ref{eqn:C66}), (\ref{eqn:D1}), and (\ref{eqn:D2}) we
obtain the remarkable result that $\widetilde{C}_{66} =0$.
 This means the nematic director relaxes to cancel out the elastic energy cost for a shear deformation.
In this limit, the nematic molecules ``wiggle'' until the deformed
structure is compatible with the boundary conditions. The stability
of the crystalline state at zero shear modulus requires higher orders
terms, like ${\cal O}(\zeta^4)$, in the elastic free energy.

To confirm these approximate analytic calculations, we numerically solve 
 Eq (\ref{eqn:dynamics}).
To examine soft elasticity, we impose a steady
shear deformation by adding an advective term on the dynamics, i.e.,
$\partial/\partial t \rightarrow \partial/\partial t + V_x \partial_x$, where 
the velocity, $V_x = \gamma y$, has a gradient in the y-axis.
We set the initial nematic director orientation such that it is along the
shear flow direction and we also shift our periodic boundaries to be consistent to the shear flow using the Lees-Edwards method~\cite{Lees72}.
We choose the following numerical parameters: $\tau =-0.03$,
$\lambda=-0.9$, and $K=1$, such that the network forms a hexagonal lattice.

The influence of the shear on the total free energy density $F_s$ is displayed
in Fig.~\ref{fig:soft1} for three values of the thermal noise strength, $\mu$.
At the lowest temperature the free energy is flat at small strains up to $\gamma t \approx 0.25$. Thus the system has the same free energy as
the undeformed state at small strains.
\begin{figure}
  \resizebox{80mm}{!}{\includegraphics{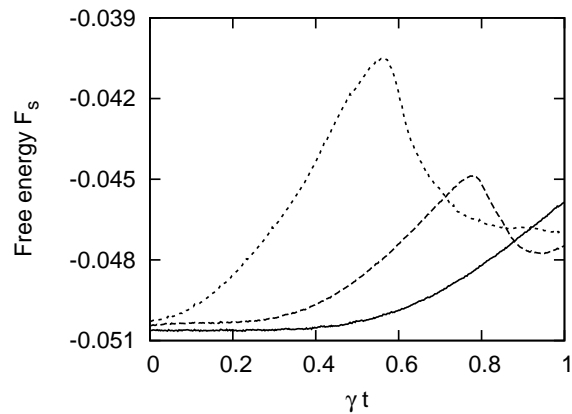}}
\caption{\label{fig:soft2}The free-energy density evolution after
commencement of a shear flow with a shear rate $\gamma = 0.002$.
The system is shown at different molecular ratio:
$\kappa =1.0$, $1.35$, and $1.7$,  for the dotted, dashed,
and solid lines respectively.($\mu=1.0$)}
\end{figure}
\begin{figure}
  \resizebox{85mm}{!}{\includegraphics{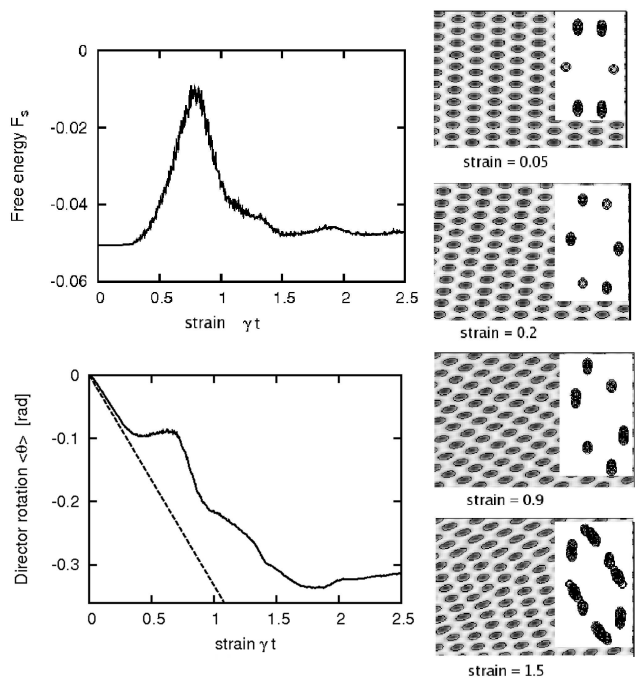}}
\caption{\label{fig:soft3}  Evolution of the free-energy density $F_s$ and the
nematic director orientation after commencement of a steady flow with shear
 a strain rate $\gamma =0.0025$.
The snapshots of the density configuration correspond to the data at the given strains.
 The dashed line
corresponds to the linear elastic theory.
($\kappa =2, \mu=1.0$)}
\end{figure}
However at higher temperatures, $F_s$ increases for all
shears as it is expected for a conventional network.
These results are consistent with Eq.~(\ref{eqn:soft}),
since the coupling constants $D_1$ and $D_2$ vanish with increasing
temperature. Thus soft elasticity vanishes with increasing temperature.
We also examined the influence of the network anisotropy on
the shear deformation and the
results are displayed in Fig.~\ref{fig:soft2}.  When $\kappa =1$ the
two fields are decoupled, and the elastic response is that of classical solid.
When $\kappa=1.7$ a soft regime is observed up to $\gamma t \approx 0.3$ as indicated by  the flat free-energy density.

Finally it is interesting to consider the configuration of the network
density $\psi(\mathbf{r})$ and the spatial average orientation of the
nematic molecules $\langle \theta(\mathbf{r})\rangle$ as
the system is being sheared.
The snapshots in Fig.~\ref{fig:soft3} show that
the network structure changes with increasing shear strain even in the 
elastic soft regime.  A uniaxial hexagonal lattice 
has a continuous set of structures~\cite{Campbell88}
which within our mean field theory have the same free energy.
The switching of $\langle \theta(\mathbf{r})\rangle$
is different from the rotational component of the shear which is
given by $\omega_{xy} =\gamma t/2$.  The initial dependence of
$\langle \theta(\mathbf{r})\rangle$ on the shear can be
calculated by minimizing the
elastic energy $E_{el}$ for a simple shear deformation applied
along the nematic director.
 This gives
\begin{equation}
\theta_{soft}(t) = (D_1+D_2)\zeta(t)/(2D_1) = \zeta(t)/(1-\kappa^2)
\label{eqn:thetasoft}
\end{equation}
in the long wavelength limit.  As shown in Fig.~\ref{fig:soft3} this
prediction works quite well.  After this soft regime the strain energy
increases until a yield occurs, e.g., for $\kappa=2$
at $\gamma t\approx 0.75$.  Interestingly
this first yield occurs without the nucleation of mobile dislocations as
would occur in a normal crystalline material.  Of course at higher strain
dislocations do eventually appear as can be seen in Fig.~\ref{fig:soft3}.
We have also made numerical calculations in the case of a static strain. 
For small strains $\langle \theta(\mathbf{r})\rangle$ relaxes to the value 
determined by Eq.~(\ref{eqn:thetasoft}) and the total free energy decays 
to the value of the undeformed state in agreement with our mean field 
predictions. We tested our numerical computations for 
$-0.03 \leq \tau \leq -0.3$ where the hexagonal phase is stable,
 and these simulations indicate that the analytic results are 
exact as $\tau \rightarrow 0_-$, with small corrections at larger $|\tau|$.
The limit $\tau \rightarrow 0_-$ is also where our sinusoidal density 
approximation is valid.
 
 In summary, a model of liquid crystal networks
was presented and shown by analytical and numerical methods to
reproduce soft elasticity as a function of temperature and
molecular shape. In addition the numerical simulations provide 
evidence of unusual non-linear yielding mechanisms which provide 
avenues of future research. 
  
 K.R.E. acknowledges the support from NSF under Grant No. DMR-0413062.
M.G. was supported by the Natural Sciences and Engineering Research Council
of Canada and by the le Fonds Qu\'eb\'ecois de la recherche
sur la nature et les
technologies.

\end{document}